\documentclass[]{aastex631}

\usepackage{cancel}

\shorttitle{Effects of gravitational radiation on the periodic FRBs}
\shortauthors{Lin et al.}
\graphicspath{{./}{figures/}}

\begin{document}

\title{Effects of Gravitational Wave Radiation of Eccentric
Neutron Star-White Dwarf Binaries on the Periodic Activity of
Fast Radio Burst Sources}

\correspondingauthor{Yi-Qing Lin;Wei-Min Gu}
\email{yqlin@xmut.edu.cn;guwm@xmu.edu.cn}

\author{Yi-Qing Lin}
\affiliation{School of Opto-electromic and Communication Engineering,
 Xiamen University of Technology, Xiamen, Fujian 361024, China}

\author{Hao-Yan Chen}
\affiliation{Department of Astronomy, Xiamen University,
Xiamen, Fujian 361005, China}

\author{Wei-Min Gu}

\affiliation{Department of Astronomy, Xiamen University,
 Xiamen, Fujian 361005, China}

\author{Tuan Yi}

\affiliation{Department of Astronomy, Xiamen University,
 Xiamen, Fujian 361005, China}

\begin{abstract}
We revisit the eccentric neutron star (NS)-white dwarf (WD) binary
model for the periodic activity of fast radio burst (FRB) sources,
by including the effects of gravitational wave (GW) radiation.
In this model, the WD fills its Roche lobe at the periastron
and mass transfer occurs from the WD to the NS.
The accreted materials can be
fragmented and arrive at the NS episodically,
resulting in multiple bursts through curvature radiation.
Consequently, the WD may be kicked away owing to
the conservation of angular momentum.
To initiate the next mass transfer,
the WD has to refill its Roche lobe through GW radiation.
In this scenario, whether the periodic activity can show up relies on
three timescales, i.e., the orbital period $P_{\rm orb}$,
the timescale $T_{\rm GW}$ for the Roche lobe to be refilled,
and the time span $T_{\rm frag}$ for all the episodic events corresponding
 to each mass transfer process.
Only when the two conditions $T_{\rm GW} \la P_{\rm orb}$ and
$T_{\rm frag} < P_{\rm orb}$ are both satisfied,
the periodic activity will manifest itself and the period should be
equal to $P_{\rm orb}$.
In this spirit, the periodic activity is more likely to show up
for relatively long periods ($P_{\rm orb} \ga$ several days).
Thus, it is reasonable that FRBs 180916 and 121102,
the only two sources having been claimed to manifest periodic activity,
both correspond to relatively long periods.
\end{abstract}


\section{Introduction} \label{sec:intro}

Fast radio bursts (FRBs) are a kind of energetic radio transients with
short durations ($\sim \rm ms$) and extremely high brightness temperatures.
The discovery of the first FRB (010724) by
\citet{2007Sci...318..777L} opens a new window in astronomy.
Recently, the number of the detected FRBs has rapidly increased
owing to the improved observational techniques and new facilities,
while the physical origin of FRBs remains a mystery
   \citep{2018PrPNP.103....1K, 2019A&ARv..27....4P,
   2019ARA&A..57..417C, 2019PhR...821....1P}.
It was found that some FRBs showed repeating behaviors
 (called repeating FRBs),
while most of them are one-off bursts
\citep[see][for a review]{2019ARA&A..57..417C}.
It is also unclear whether repeating FRBs and one-off FRBs have
the same physical origin \citep{2018ApJ...854L..12P, 2019MNRAS.484.5500C}.

The Canadian Hydrogen Intensity Mapping Experiment Fast Radio Burst
Project (CHIME/FRB) \citep{2019Natur.566..235C} has detected a
large number of FRBs and the first source with periodic activity
(FRB 180916) has been reported \citep{2020Natur.582..351C}.
Since the report of the periodic activities,
the physical origin has been widely investigated.
Several kinds of models have been proposed to explain the periodic
activities \citep[for a review, see][]{2021SCPMA..6449501X}.
First, FRBs occur in a binary system containing a stellar
compact object (a neutron star (NS) or a black hole (BH)),
where the observed period corresponds to the orbital period
\citep{2016ApJ...829...27D, 2017ApJ...836L..32Z, 2020ApJ...895L...1D,
2020MNRAS.497.1543G, 2021arXiv210204264K, 2021arXiv210206796D,
2021arXiv210514480W, 2021arXiv210800350L,2021arXiv210304165G,2021MNRAS.508.2079V}.
Second, an extremely slow rotation of an NS causes
the ultra-long rotational period of the bursting source
\citep{2020MNRAS.496.3390B, 2021ApJ...917....2X}.
Third, FRBs are emitted from the NS magnetosphere due to internal
triggers or external triggers, and the long-lived precession
of the emitting region causes the observed periodic activities
\citep{2020ApJ...895L..30L, 2020RAA....20..142T,
2020ApJ...893L..31Y, 2021ApJ...909L..25L,2021ApJ...917...13S}.

To date, FRBs 180916 and 121102 are the only two FRBs having been claimed
to have periodic activity.
FRB 180916 exhibits an activity period of
$16.35\pm 0.18 $ days with an activity window
 about 5 days \citep{2020Natur.582..351C}.
Observationally, this source is not always
active in all the predicted activity windows.
FRB 121102 shows a possible 157-day
periodic behavior and an active window $\sim 100$ days
\citep{2020MNRAS.495.3551R, 2021MNRAS.500..448C}.
On the other hand, for some repeaters with tens or even more than
a thousand observed bursts,
no periodic activity was found by time-series analyses
\citep{2021arXiv211007418N,2021arXiv211111764X}.

In this paper, based on the NS-white dwarf (NS-WD) binary
model with an eccentric orbit \citep{2020MNRAS.497.1543G},
we will investigate the effects of GW
radiation on the periodic activity of FRB sources.
We will manage to
understand:
(1) why the sources are not active in some predicted windows,
(2) why no periodic activity was discovered for some repeaters with
adequate burst-observations,
 and (3) the observed periods for the recurrence of activity (16.35
or 157 days) are significantly longer than the orbital periods of
contact NS-WD binaries with moderate eccentricities (tens of minutes).
The remainder of this paper is organized as follows.
The NS-WD binary model with the GW radiation and
the corresponding equations are described in Section~\ref{sec:binary model}.
The results are shown in Section~\ref{sec:results}.
Our conclusions and discussion are presented in Section~\ref{sec:conclusion}.

\section{NS-WD binary model with gravitational radiation}
\label{sec:binary model}

\citet{2020MNRAS.497.1543G} proposed a compact binary
model with an eccentric orbit to explain the periodic activity of
repeating FRB sources.
The model includes a magnetic WD and an
 NS with strong dipolar magnetic fields.
When the WD fills its Roche lobe at the periastron, mass transfer occurs
from the WD to the NS, and the accreted and fragmented materials
can produce multiple FRBs when traversing along the NS's magnetic field.
 In this scenario, the period for the recurrence of activity should
be equivalent to the orbital period $P_{\rm orb}$.
Owing to the conservation of angular momentum, the WD may be kicked away
after a mass-transfer process for $q<2/3$ \citep{2007RSPTA.365.1277K},
where $q$ is the mass ratio defined as $q\equiv M_2/M_1$.
Here, $M_1$ and $M_2$ are the NS and WD masses, respectively.
As proposed by \citet{2016ApJ...823L..28G}, the WD can refill its Roche lobe
through the GW radiation. Thus, it is essential to compare the two
key timescales, i.e., $P_{\rm orb}$ and $T_{\rm GW}$, where
$T_{\rm GW}$ is the timescale for the Roche lobe to be refilled.

\begin{figure}
\includegraphics[width=15cm]{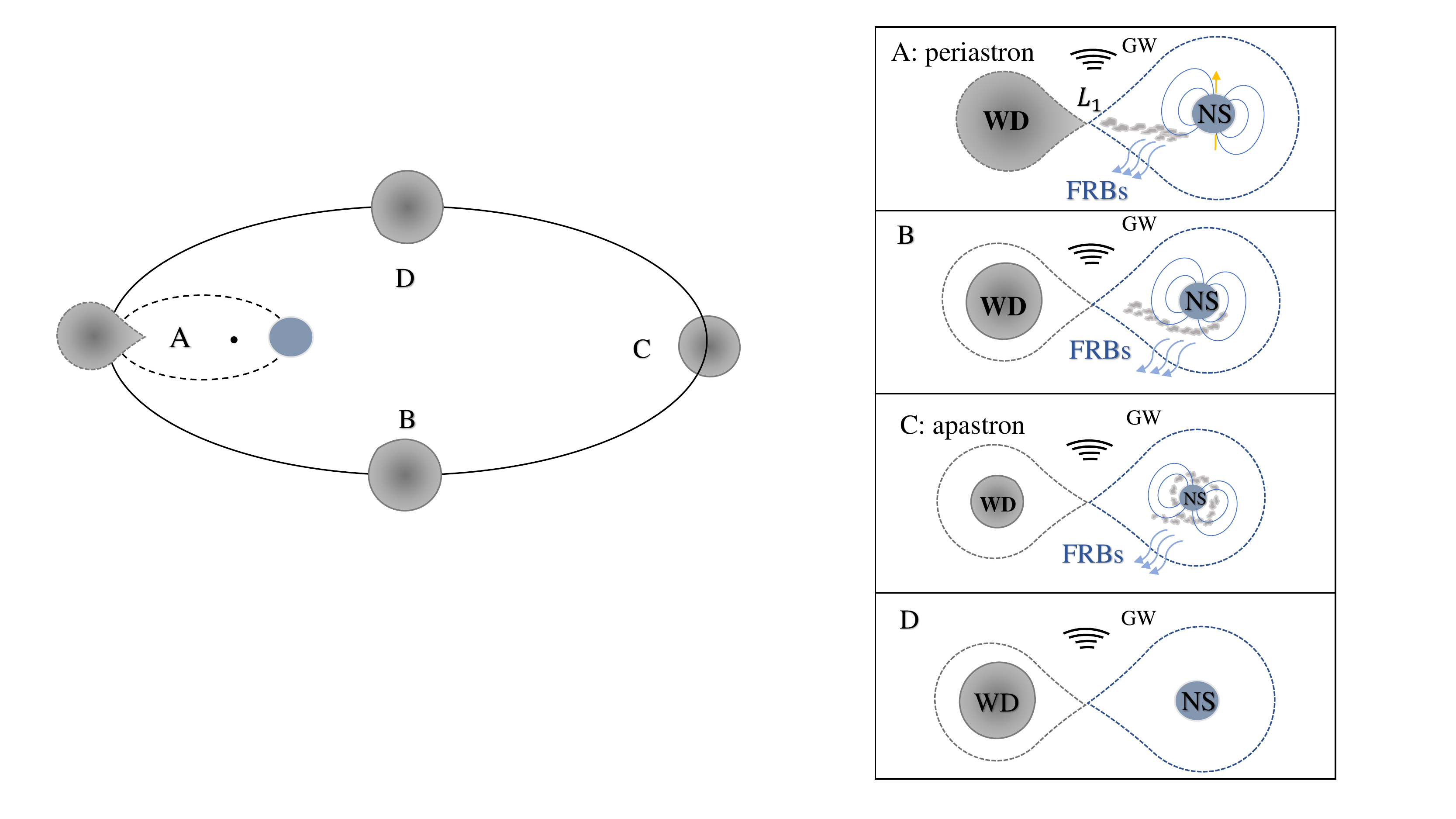}
\centering
\setlength{\abovecaptionskip}{0.5cm}
\caption{
Illustration of the eccentric NS-WD binary model.
Panel (A): the WD fills its Roche lobe at the periastron, and mass transfer occurs
from the WD to the NS through the $L_1$ point; Panels (B, C, and D):
the WD travels to the other phases of the eccentric orbit,
where the Roche lobe is not filled and the mass transfer is interrupted.
The fragmented materials trigger multiple FRBs.
The WD refills its Roche lobe at the next periastron
 passage, due to the decrease of orbital separation through GW radiation,
 which takes a timescale of $T_{\rm GW}$.
}
\label{f1}
\end{figure}

Figure \ref{f1} is a cartoon animation of our model that describes an NS-WD
system with an eccentric orbit. Similar to \citet{2020MNRAS.497.1543G},
a mass transfer occurs from the WD to the NS when the WD fills
its Roche lobe near the periastron (panel A).
Owing to the viscous processes, the accreted WD material can be
fragmented and arrive onto the NS episodically.
However, at other positions on the eccentric orbit,
no mass transfer is supplied since
the Roche lobe is not filled, as shown in panels (B, C, and D)
of Figure \ref{f1}. The fragmented materials travel along the
magnetic field lines (on to the NS) and results in
multiple bursts by the curvature radiation.

We introduce two timescales during the accretion process.
The first one is the accretion timescale $T_{\rm acc}$,
which is the timescale for the accreted materials to transfer from
the inner Lagrangian point to the surface of the NS.
Since the accreted materials have angular momentum and therefore
viscous processes are necessary, $T_{\rm acc}$ is normally
longer than the orbital period $P_{\rm orb}$.
In our model, for the mass transfer process near the periastron
in different cycles, the timescale $T_{\rm acc}$ is regarded to be a
certain fixed value.
The second one is the timescale $T_{\rm frag}$, which denotes
the time span from the first piece of materials arrival at the surface
of the NS to the arrival time of the last piece. $T_{\rm frag}$ corresponds
to the activity window and can be either shorter or longer than $P_{\rm orb}$.
In this work, we will focus on $T_{\rm frag}$ rather than $T_{\rm acc}$,
since the latter works as a unified time-lag and may not have significant
effects on the observations. On the contrary,
$T_{\rm frag}$ is an important parameter that can affect the
periodicity of the observed FRBs, which will be discussed
in Section~\ref{sec:results}.

On the other hand, the WD may be kicked away after a mass-transfer process,
and thus widen the orbital separation. The GW radiation
is an essential mechanism to reduce the
orbital separation and enable the WD to refill its Roche lobe,
initiating to the next mass transfer event.
Thus, the time interval between two adjacent mass-transfer process $T_{\rm GW}$
is a key timescale, which may determine whether the periodic activity
can show up.
Obviously, $T_{\rm GW} \la P_{\rm orb}$ is a necessary condition for
the periodic activity to appear.
On the contrary, for $T_{\rm GW} \gg P_{\rm orb}$,
since $T_{\rm GW}$ also varies with different amount of mass transfer near
the periastron (Equation~(\ref{e9})),
the periodicity will be hard to be discovered.

The dynamic equation of a binary is given by
\begin{equation}
{G(M_1+M_2)\over a^3}={4\pi^2\over P_{\rm orb}^2},
\label{e1}
\end{equation}
where $G$ is the gravitational constant,
and $a$ is the semi-major axis of the eccentric orbit. The effective
Roche lobe radius for the WD $R_{\rm L2}$ at the periastron can be simply
described as \citep{1971ARA&A...9..183P}
 \begin{equation}
{R_{\rm L2} \over a(1-e)}=0.462\left({M_2\over{M_1+M_2}}\right)^{1/3},
\label{e2}
\end{equation}
where $e$ is the eccentricity of the orbit. The WD radius $R_{\rm WD}$
depends only on its mass and composition \citep{1997MNRAS.291..732T}.
For $M_2\la 0.2\ M_\sun$, it is convenient to simplify
Equation (17) of \citet{1997MNRAS.291..732T} as
\begin{equation}
R_{\rm WD}=0.0115R_{\sun}({M_{\rm CH}/M_2})^{1/3},
\label{e3}
\end{equation}
where $R_{\sun}$ is the solar radius, and $M_{\rm CH}=1.44\ M_{\odot}$ is
the Chandrasekhar mass limit.
The orbital angular momentum $J$ of a binary system in an eccentric orbit
can be expressed as \citep{1964PhRv..136.1224P}
\begin{equation}
J = M_1M_2\left[{Ga(1-e^2)\over M_1+M_2}\right]^{1/2}.
\label{e4}
\end{equation}

When the WD fills its Roche lobe at periastron, the mass transfer occurs.
With the assumption of the orbital angular momentum conservation
($\Delta J = 0$) and the assumption that the eccentricity $e$ does not
change during the whole process, the variation $\Delta a$
due to the mass transfer can be derived from Equation (\ref{e4}):
\begin{equation}
{\Delta a\over a}=2(1-q){\Delta M_2\over M_2},
\label{e5}
\end{equation}
where the positive $\Delta M_2$ is the transferred mass near the periastron.
On the other hand, the WD radius will expand with decreasing mass.
By Equations (\ref{e2}) and (\ref{e3}), we obtain the required variation of semi-major
axis $\Delta a_*$ for the WD just to refill its Roche lobe
($\Delta R_{\rm WD} = \Delta R_{\rm L2}$):
\begin{equation}
{\Delta a_*\over a} = {2\over 3}{\Delta M_2\over M_2}.
\label{e6}
\end{equation}
The critical condition $\Delta a = \Delta a_*$ results in a critical
mass ratio $q = 2/3$, as shown by \citet{2007RSPTA.365.1277K}.
Thus, for $q > 2/3$, we have $\Delta a < \Delta a_*$ and
$\Delta R_{\rm WD} > \Delta R_{\rm L2}$, which indicates that
the expansion of the WD radius is more rapid than
that of its Roche lobe \citep[e.g.][]{2018MNRAS.475L.101D}.
Thus, the Roche lobe will be over-filled at periastron
for each time. However, our study focuses on $M_2 \la 0.2 \ M_{\odot}$ and
therefore the values of $q$ is far below $2/3$ in the NS-WD model.
In other words, we only have $\Delta a > \Delta a_*$.
In this case, the Roche lobe can be refilled only when the GW radiation
is taken into account.

For the binary with an eccentric orbit, the variation of $a$ due to
the GW radiation can be written as \citep{1964PhRv..136.1224P}
\begin{equation}
{da\over dt}=-{64\over 5}{G^3M_1M_2(M_1+M_2)\over c^5a^3(1-e^2)^{7/2}}
\left( 1+{73\over 24}e^2+{37\over 96}e^4 \right).
\label{e7}
\end{equation}
The time interval $T_{\rm GW}$ between two adjacent mass-transfer process
can be derived as
\begin{equation}
T_{\rm GW} = - {\Delta a - \Delta a_* \over{da/dt}}.
\label{e8}
\end{equation}
Thus, an analytic relation between $T_{\rm GW}$ and $\Delta M_2$
can be expressed as
\begin{equation}
T_{\rm GW} = {5c^5a^4({2\over 3}-q)(1+q)^2\over 32qG^3(M_1+M_2)^3}
{(1-e^2)^{7/2}\over 1+{73\over 24}e^2+{37\over 96}e^4}\cdot
{\Delta M_2\over M_2}.
\label{e9}
\end{equation}
Equation (\ref{e9}) enables us to calculate $T_{\rm GW}$ once $M_1$, $M_2$, $e$,
and $\Delta M_2$ are given. In our model, we adopt a typical value
$M_1 = 1.4 \ M_{\sun}$ for the NS.

\section{Results}
\label{sec:results}

According to the assumption that the WD fills its Roche lobe at periastron,
i.e., $R_{\rm L2}=R_{\rm WD}$, we can obtain the values of $P_{\rm orb}$ by
Equations (\ref{e1})-(\ref{e3}) and $T_{\rm GW}$ by Equation (\ref{e9}) once $M_2$, $e$, and
$\Delta M_2$ are given. $T_{\rm GW}$ and $ P_{\rm orb}$ are two key
timescales which may determine whether the periodic activity can show up.
We use the ratio $T_{\rm GW}/P_{\rm orb}$ to show the effects of
the GW radiation on the periodic activity.
\begin{figure}[h]
\centering
\setlength{\abovecaptionskip}{-1cm}
\includegraphics[width=120mm,height=150mm]{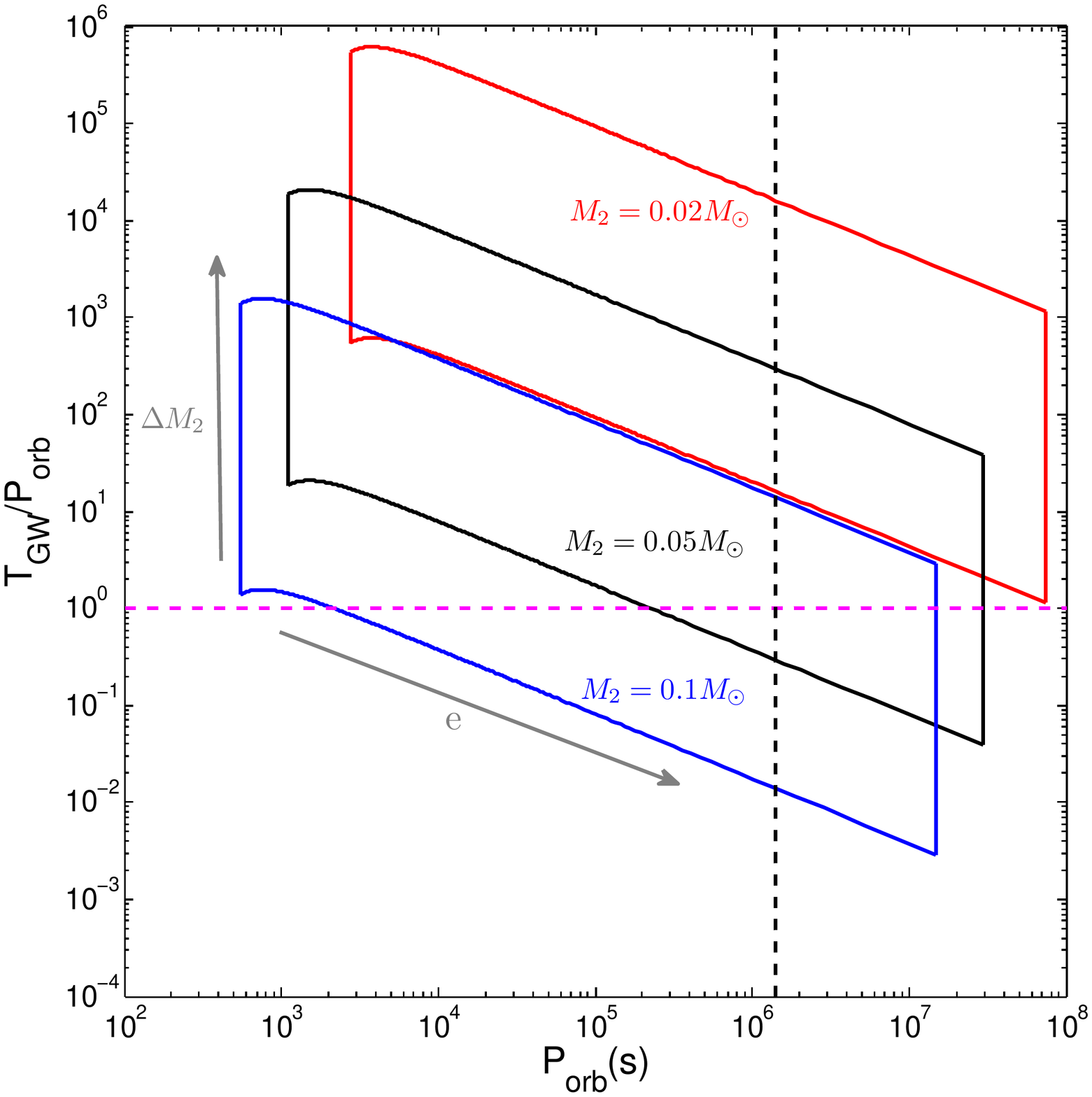}
\caption{
The $T_{\rm GW}/P_{\rm orb}$-$P_{\rm orb}$ diagram with
eccentricities $e = 0.1-0.999$ and the mass transfer
$\Delta M_2 = 10^{-12}-10^{-9} \ M_{\odot}$ for three masses
$M_2 = 0.02 \ M_{\odot}$ (red solid lines),
$0.05 \ M_{\odot}$ (black solid lines),
and $0.1 \ M_{\odot}$ (blue solid lines).
The pink horizontal dashed
line shows the critical relation $T_{\rm GW} = P_{\rm orb}$, the
black vertical dashed line represents the reported 16.35-day
period of FRB 180916, and the two gray arrows
indicate the increasing direction of $\Delta M_2$ and $e$, respectively.
}
\label{Fig:fig2}
\end{figure}

Hydrodynamic simulation of NS-WD binaries showed that only the systems
with $M_2 < 0.2\ M_{\odot}$ can undergo stable mass transfer,
while the systems with $M_2>0.2\ M_{\odot}$ experience unstable
mass transfer, which may lead to tidal disruption of
the WD \citep{2017MNRAS.467.3556B}. Thus, a stable NS-WD binary likely
to have a mass range $0.01 M_{\odot} < M_2 < 0.2\ M_{\odot}$
\citep{2020MNRAS.497.1543G}.
With $M_1 = 1.4\ M_{\sun}$ and $0.01\ M_{\odot} < M_2 < 0.2\ M_{\odot}$,
i.e. $q < 1/7$, the WD may be kicked away after a mass-transfer process
\citep{2007RSPTA.365.1277K}.

Figure \ref{Fig:fig2} shows the $T_{\rm GW}/P_{\rm orb}$ - $P_{\rm orb}$
diagram with a pair of physically plausible ranges
$\Delta M_2 = 10^{-12}-10^{-9} \ M_{\odot}$ \citep[e.g.,][]{2016ApJ...823L..28G}
and $e=0.1-0.999$ for
three masses $M_2 = 0.02\ M_{\odot}$ (red solid lines),
$M_2 = 0.05 \ M_{\odot}$ (black solid lines),
and $M_2 = 0.1 \ M_{\odot}$ (blue solid lines). The pink horizontal dashed
line corresponds to $T_{\rm GW} = P_{\rm orb}$, and the black vertical
dashed line represents the reported 16.35-day period of FRB 180916.
It is seen from Figure \ref{Fig:fig2} that, the values of $T_{\rm GW}/P_{\rm orb}$
fall into the quadrilateral regions.
As mentioned in Section~\ref{sec:binary model}, $T_{\rm GW} \la P_{\rm orb}$ is a necessary condition for the periodic activity to appear. In addition,
the period should be equivalent to $P_{\rm orb}$.
On the contrary, for $T_{\rm GW} \gg P_{\rm orb}$,
since $T_{\rm GW}$ also varies with different $\Delta M_2$ near
the periastron, the periodicity will be hard to be discovered.
In this spirit, the region below the pink dashed line corresponds
to the occurrence of periodic activity. Taking the results for
$M_2 = 0.05 \ M_{\odot}$  as an example, it is seen from Figure \ref{Fig:fig2} that,
even though the orbital period $P_{\rm orb}$ can cover a wide range
($10^3$ to $10^7$ seconds), only the systems with $P_{\rm orb} \ga 10^6$
seconds may present periodicity.
To date, only two sources, FRBs 180916 and 121102, have been claimed
to show periodic activity. According to our model, it is quite reasonable
that relatively long periods ($\ga$ several days) show up both in
these two sources.

Moreover, since $T_{\rm GW}$ is proportional to $\Delta M_2$ (Equation (\ref{e9})),
the condition $T_{\rm GW} \la P_{\rm orb}$ may be
satisfied only for some cycles with relatively low values of $\Delta M_2$,
and not be satisfied for some other cycles with relatively high values of
$\Delta M_2$.
In this scenario, we can understand why FRB 180916 has periodic activity
but is not always active in all the predicted windows.

Another unresolved problem is that, for some repeaters with tens or even more than a
thousand observed bursts, such as FRBs 190520B \citep{2021arXiv211007418N}
and 201124A \citep{2021arXiv211111764X},
no periodic activity was found based on time-series analyses.
In our opinion, the reason is related to $T_{\rm frag}$,
which denotes the time span
from the first piece of materials arrival at the NS surface to the
arrival time of the last piece, as described in Section~\ref{sec:binary model}.
Since $T_{\rm frag}$ is quite an uncertain timescale in our model,
it is regarded as a free parameter.
For $T_{\rm frag} < P_{\rm orb}$, it is clear that $T_{\rm frag}$
will work as a relatively narrow activity window in each cycle.
In order to interpret the observations of FRBs 180916 and 121102
by our model, $T_{\rm frag}$ is required to be around 5 and 100 days,
respectively.
For $T_{\rm frag} \ga P_{\rm orb}$,
however, the periodic activity will be concealed and difficult to be revealed.
Thus, the frequently repeating sources without periodic activity may
correspond to $T_{\rm frag} \ga P_{\rm orb}$.

\section{Conclusions and Discussion}
\label{sec:conclusion}

In this paper, we have revisited the eccentric NS-WD binary model
by including the effects of GW radiation on the periodic activity of FRB sources.
We have shown that, even though our model
indicates that the cycle time of the burst activity should be equivalent
to the orbital period, whether the periodic activity can be present
is related to the three timescales,
i.e., the orbital period $P_{\rm orb}$,
the timescale $T_{\rm GW}$ for the WD to refill its Roche lobe through
the GW radiation,
and the time span $T_{\rm frag}$ related to
the different arrival time for the fragmented materials.
Our analyses indicate that:
(1) Only when the conditions $T_{\rm GW} \la P_{\rm orb}$
and $T_{\rm frag} < P_{\rm orb}$ are both satisfied, the periodic activity
will appear. Otherwise,
for either $T_{\rm GW} \gg P_{\rm orb}$ or
$T_{\rm frag} \ga P_{\rm orb}$, the periodicity will be hard to discovered.
(2) Since $T_{\rm GW}$ is proportional to $\Delta M_2$,
it is understandable that FRB 180916
is not always active in all the predicted activity windows.
(3) FRBs with relatively long periods ($\ga$ several days) are more
likely to show up.
Thus, it is reasonable that the only two sources having been claimed to
have periodic activity, i.e. FRBs 180916 and 121102, both correspond
to relatively long periods.

According to our model, a large eccentricity is required for the periodic
activity to show up. In our opinion, there are two possibilities for
the formation of an NS-WD binary with a large eccentricity.
One is the evolution result of a binary channel by the natal kick
of the supernova explosion.
Since both gravitational waves and accretion tend to circularise the orbit,
a highly eccentric orbit in this scenario should correspond to a young NS.
The other possibility is a WD captured by an NS,
where an old NS is quite possible. Whether the second possibility plays an
important role may be inferred from recent statistical analysis.
The CHIME/FRB Collaboration released their first FRB
catalogue which includes 536 FRB events \citep{2021ApJS..257...59A}.
This uniform large sample allows a better statistical analysis of
the FRB population. \citet{2022ApJ...924L..14Z} tested
the new CHIME sample against the star-formation rate density, cosmic
stellar-mass density, and delayed models. They concluded that the CHIME FRB
population do not track the star formation history of the Universe,
and their results indicate the old population as the origin of FRBs.
In addition, \citet{2022MNRAS.tmp..120H} showed that old populations such as
old NSs and BHs are more likely progenitors of non-repeating FRBs.
Thus, with regards to these statistical results, for our model,
it is required that the capture mechanism should have significant
contributions to the formation of the eccentric NS-WD binaries.

Notably, the periodic activity behavior may be much more complex
and diverse than our current understandings.
Recently, \citet{2021ApJ...911L...3P} reported that
the burst activity of FRB 180916 is systematically delayed toward lower
frequencies by about three days (0.2 cycles) from 600 to 150~MHz.
They also discussed a possible link between the frequency dependence
of the observed activity and a radius-to-frequency mapping effect
for various models.
Such an issue is beyond the scope of the present paper.

It should be noted that, outflows are not taken into account in this work.
Outflows may carry angular momentum and escape from the binary system,
which corresponds to $\Delta J < 0$ rather than $\Delta J = 0$.
\citet{2018MNRAS.475L.101D} investigated such an issue and showed that
a violent mass transfer may occur even for $q < 2/3$, such as a BH-WD
system. It is easy to understand that, if outflows carry away significant
angular momentum, the timescale $T_{\rm GW}$ can be greatly
shortened, which is
helpful to satisfy the condition $T_{\rm GW} \la P_{\rm orb}$, and therefore
the periodic activity is more likely to show up.

\begin{acknowledgments}
We thank the referee for helpful suggestions that improved the manuscript.
This work was supported by the National Natural Science Foundation of China
under grants 11925301, 12033006, and 11573023.
We acknowledge the science research grants from the China Manned Space Project
with NO. CMS-CSST-2021-B07 and CMS-CSST-2021-B11, and acknowledge support
from the China Postdoctoral Science Foundation under grant 2021M702742.

\end{acknowledgments}

\bibliography{sample631}{}
\bibliographystyle{aasjournal}

\end{document}